\documentclass[%
reprint,
superscriptaddress,
amsmath,amssymb,
aps,prl
]{revtex4-1}

\usepackage{graphicx}
\usepackage{pstool}
\usepackage{psfrag}
\usepackage{multirow}
\graphicspath{{../Pictures/}}
\usepackage{dcolumn}
\usepackage{bm}
\usepackage[percent]{overpic}

\begin{document}
	
	
	\title{Optical coupling between resonant dielectric nanoparticles and dielectric waveguides probed by third harmonic generation microscopy}
	
	
	\author{Kirill I. Okhlopkov}
	\affiliation{Faculty of Physics, Lomonosov Moscow State University, 19991 Moscow, Russia}
	
	\author{Alexander A. Ezhov}
	\affiliation{Faculty of Physics, Lomonosov Moscow State University, 19991 Moscow, Russia}
	
	\author{Pavel Shafirin}
	\affiliation{Faculty of Physics, Lomonosov Moscow State University, 19991 Moscow, Russia}
	
	\author{Nikolay~A.~Orlikovsky}
	\affiliation{Bauman Moscow State Technical University, 105005 Moscow, Russia}

	\author{Maxim R. Shcherbakov}
	\affiliation{Faculty of Physics, Lomonosov Moscow State University, 19991 Moscow, Russia}
	\affiliation{School of Applied and Engineering Physics, Cornell University, Ithaca, 14853 NY, USA}

	\author{Andrey A. Fedyanin}
	\email[]{fedyanin@nanolab.phys.msu.ru}
	\affiliation{Faculty of Physics, Lomonosov Moscow State University, 19991 Moscow, Russia}

	
	\date{\today}

\begin{abstract}
Localized electromagnetic modes and negligible Ohmic losses dictate the growing interest in subwavelength all-dielectric nanoparticles.
Although an exhaustive volume of study dealt with interaction of all-dielectric nanostructures with free-space electromagnetic fields, their performance as integrated photonics elements remains untackled.	
We present an experimental study of optical coupling between a resonant subwavelength silicon nanodisk and a non-resonant silicon waveguide, as probed by third harmonic generation microscopy. By placing the nanodisks at different distances from the waveguide, we observe third harmonic intensity modulation by a factor of up to 4.5. 
This modulation is assigned to changes in the local field enhancement within the nanodisks caused by their coupling to the waveguides and subsequent modulation of their magnetic-type resonances. Interestingly, although the waveguide presents an additional loss channel for the nanodisk, we observe an increase in the local field strength within the nanodisk, as verified by rigorous full-wave simulations. This work makes a step toward integration of all-dielectric nanoparticles on photonic chips.
\end{abstract}


\maketitle

	\section{Introduction}
	
Nanophotonics deals with optical properties of nanoscale materials, of which nanoparticles (NPs) play a central role, for they exhibit strong shape-dependent resonances  utilized in light harvesting \cite{Atwater2010}, medicine and biological applications \cite{Jain2008}.
While the shape of NPs in many ways determines their optical properties \cite{Jain2006}, it is important to disclose new degrees of freedom to control their resonances. Resonances of individual NPs can be affected by changing their environment \cite{Kelly2003} of by bringing them in the near-field vicinity of other resonant systems \cite{Rechberger2003}. Optical coupling between closely spaced NPs brings about hybridization of modes, giving additional means to shape their spectra and local fields \cite{Halas2011}.
This property was extensively used to demonstrate a plasmonic analog of electromagnetically induced transparency \cite{Liu2011}, desiging plasmonic waveguides \cite{Maier2003} and so-called plasmon rulers \cite{Jain2007,Teperik2013,Shcherbakov2015}.
However, many applications, including nanophotonic circuitry, impose severe restrictions to levels of nonradiative losses in materials, for which plasmonic NPs do not always qualify.

A new paradigm in nanophotonics has recently emerged, where metallic NPs are challenged by those made of high-index dielectrics \cite{Kuznetsov2016a}. Apart from studies on single-particle scattering \cite{Kuznetsov2012,Evlyukhin2012}, optical properties of dielectric NP formations in dimers \cite{Bakker2015}, trimers \cite{Shcherbakov2015}, and higher-order oligomers \cite{Hopkins2013,Shorokhov2016} have been explored, with much stress put on the fundamental mode of NPs, the magnetic dipolar (MD) mode.
Low Ohmic losses of all-dielectric NPs make them also appealing for nonlinear-optical applications, where high optical powers are a necessity \cite{Shcherbakov2014,Yang2015b,Liu2016b}.
Finally, and in contrast to plasmonic NPs, all-dielectric NPs can be fabricated using CMOS-compatible materials, most importantly silicon \cite{Staude2017}, making them perfect candidates for integrated photonics devices. Although plasmonic nanoparticles coupled to the waveguides have been studied extensively \cite{Sidiropoulos2014,Vercruysse2017a}, their dielectric counterparts have not been considered as integrated photonics elements yet.
	
	In this Article, we observe optical coupling between single subwavelength silicon nanodisks and the core element of integrated nanophotonics---a silicon waveguide. Subwavelength nanodisks are designed to exhibit magnetic dipolar resonances in the near-infrared, which are experimentally shown to increase the third-harmonic up-conversion efficiency by a factor of up to 17 with respect to bulk silicon.  Placing nanodisks within a subwavelength distance from waveguides, we observe strong modification of the local fields within the nanodisk, which results in a modulation of their nonlinear-optical response by a factor of up to 4.5. Counter-intuitively, the non-resonant waveguides can act constructively, enhancing the nonlinear response of the nanodisks, which is supported by full-wave simulations of the local fields within the nanodisks. To our knowledge, our findings present the first study of all-dielectric particles interacting with an integrated waveguide, opening new ways of tailoring their response, both in linear and nonlinear regimes.
	
%
	
\section{Results and Discussion}
	
\textbf{Sample design and fabrication} Among many shapes and designs for all-dielectric NPs that possess localized MD resonances, a nanodisk is the most well-understood geometry, from those accessible by lithography. Silicon was chosen as a material for proposed structures for its CMOS-compatibility, high refractive index, and large value of third-order susceptibility, which plays a major role in our experiments. The proposed ``silicon nanodisk---waveguide'' pairs were fabricated out of a silicon-on-insulator wafer by electron-beam lithography and reactive-ion etching. 
The thicknesses of both the device silicon and the buried SiO$_2$ layers were 280~nm. Nine combinations of silicon nanodisk diameters ($D$) and the gap sizes between a disk and a waveguide ($L$) were fabricated, as given in  Fig.1(a); here, $D=380$, 430 and 480~nm, and $L=90$, 185 and 315~nm. The purpose of varying the gap size is to modify the optical coupling between nanodisks and waveguides, whereas varying the nanodisk diameter gives us flexibility in tuning the spectral position of the MD mode. The existence of a resonance at a wavelength of 1.53~$\mu$m in the nanodisk is demonstrated numerically for $D=480$~nm by observing resonant enhancement of local electric (Fig.1(c)) and magnetic (Fig.1(d)) fields, which reveal an $x$-oriented MD-type field distribution. For smaller diameters, the central wavelength of the resonance was determined to be blue-shifted by approximately 90 and 170~nm with respect to the pump wavelength; see supporting Information for deatails. 
For the waveguide to be suitable for a wavelength of 1.53~$\mu$m and to present a potential in-/out-coupling channel, its width was chosen to be 445~nm. A scanning electron microscope image of a typical dimer with a gap size of 185 nm and with a nanodisk diameter of 480 nm is shown in Fig.1(b).
%
%

	\begin{figure}
		\begin{overpic}[width=1\linewidth]{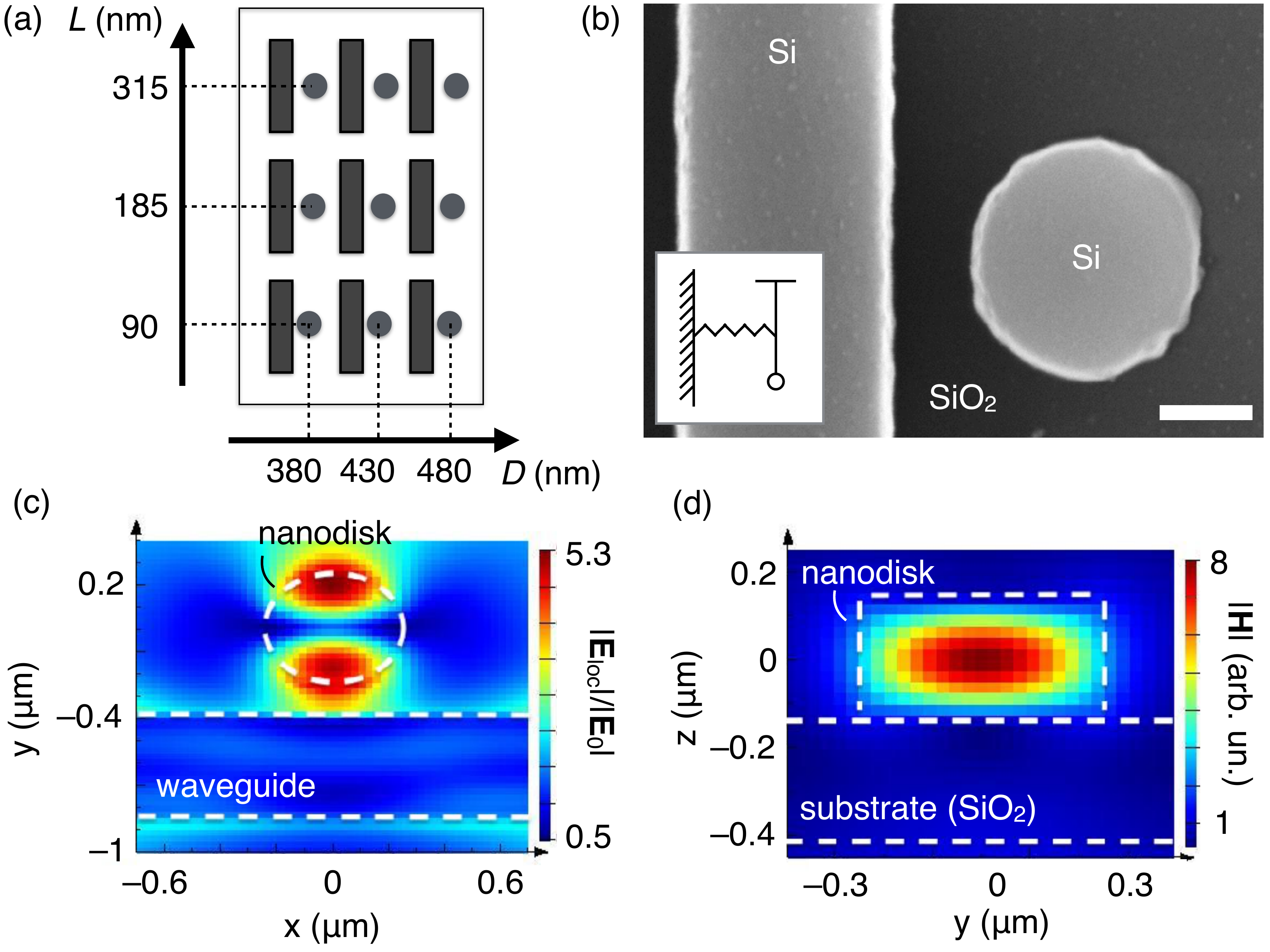}
		\end{overpic}
		\caption{\label{fig0} {(a) The set of nanodisk diameters ($D$) and gaps sizes between nanodisks and waveguides ($L$) that was chosen for fabrication. (b) A scanning electron micrograph of the sample with parameters $D = 480$~nm and $L=185$~nm. The scale bar is 200~nm. (c) Calculated distribution of normalized electric field strength in the middle $z$-section of the waveguide and nanodisk. (d) Calculated distribution of normalized magnetic field magnitude in the plane normal to substrate through the center of nanodisk. Both maps are given for  $D=480$~nm, $L=150$~nm and an excitation wavelength of $1.53$~$\mu$m. The white dashed lines indicate the boundaries of structures.} 
		}
	\end{figure}
	
	\textbf{Nonlinear microscopy} When an optical system is excited at its resonant wavelength, electromagnetic fields within the system may be much higher than those in the incoming wave. 
Optical coupling is known to disturb resonance conditions in nanoparticles, affecting the structure of the local fields, which may be, in turn, probed by measuring the nonlinear optical response \cite{Teperik2013}. Here, we probe optical coupling by illuminating the nanodisks with tightly focused femtosecond laser pulses and collecting scattered third harmonic signal generated by the nonlinear polarization $\mathbf{P}^{(3)}=\varepsilon_0\hat{\chi}^{(3)}_{\rm Si}\vdots\mathbf{E}_\omega(\mathbf{r})\mathbf{E}_\omega(\mathbf{r})\mathbf{E}_\omega(\mathbf{r})$ in silicon parts of the structure. We can safely neglect THG from the buried oxide, as the nonlinear susceptibility of SiO$_2$ is orders of magnitude smaller than components of $\hat{\chi}^{(3)}_{\rm Si}$. In our setup, the pump pulses came from an Er-ion-based fiber oscillator with a carrier wavelength of $\lambda = 1.53$~$\mu$m, as shown in Fig.2(a). The pulses produced peak intensities on the order of 10--30~GW/cm$^2$ after being focused at the sample to a waist of 3~$\mu$m in diameter.
The forward-emitted TH signal was collected by a scanning confocal microscope, which was capable of mapping the intensity of the THG over the confocal image of the sample that was taken simultaneously. 
For further measurement details, refer to Methods section.	
	
	\begin{figure*}
		\begin{overpic}[width=0.85\linewidth]{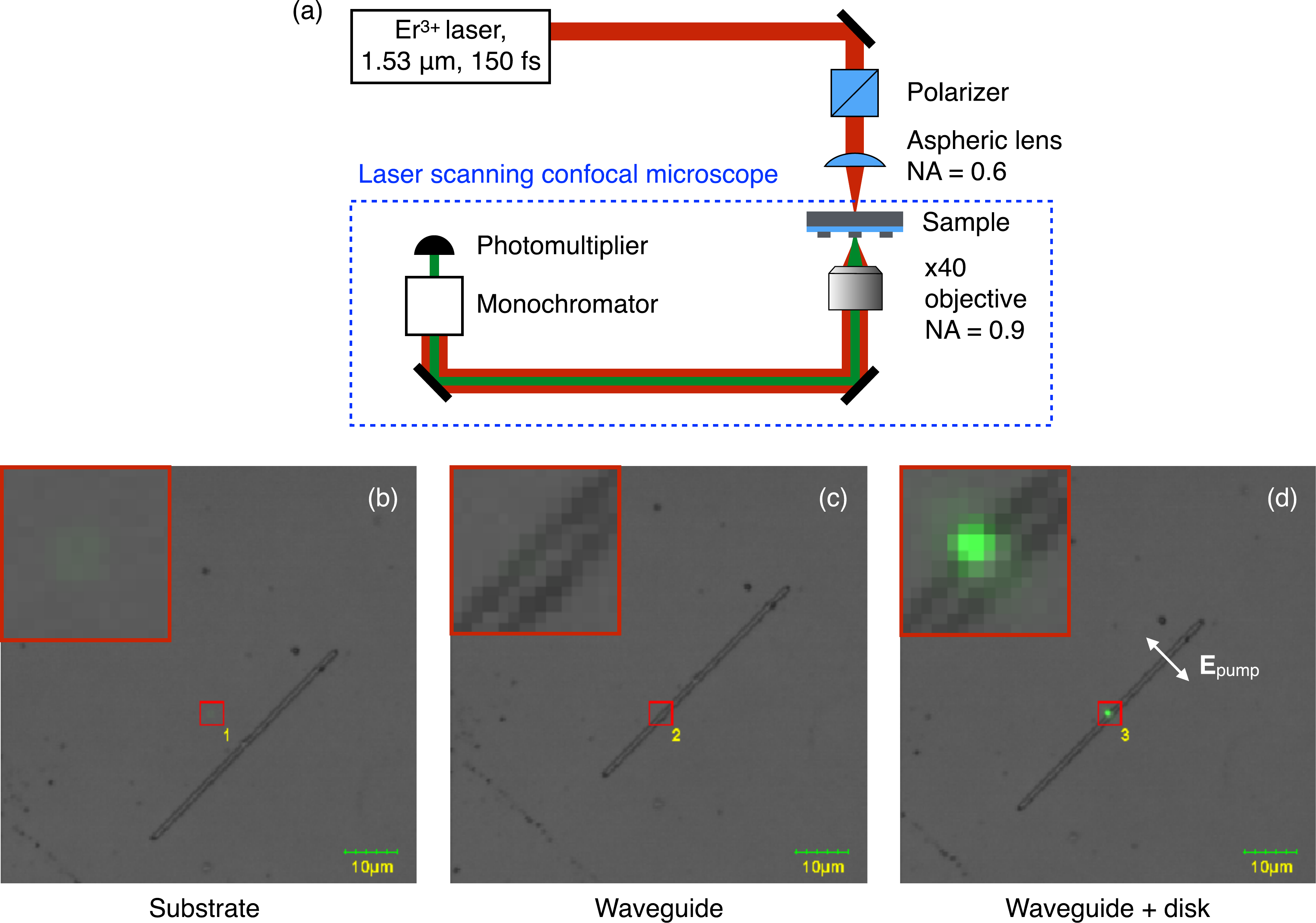}
		\end{overpic}
		\caption{\label{fig0} { 
		(a) Third harmonic generation microscopy setup. 
		(b,c,d) Third harmonic generation maps (green) laid over the confocal images of the sample (gray). Small red squares circumscribe the pump beam and represent the integration area used to measure THG signals. Three cases are shown: (b) the pump illuminates the substrate, THG is barely seen; (c) the pump illuminates the waveguide, no THG is observed; (d) the pump illuminates the waveguide and the disk, THG is bright. Insets are close-ups on the red squares. } 
		}
	\end{figure*}

	In accord with previous findings \cite{Shcherbakov2014}, we observe that exciting the MD mode of the nanodisks significantly enhances THG. Shown in Fig.2(b,c,d) are three snapshots of a ``waveguide--disk'' pair with the small red squares circumscribing the position of the pump, which was polarized perpendicularly to the waveguide. In these images, the nanodisk is situated in the middle of the waveguide and cannot be resolved due to its subwavelength size. The grey-scale confocal images are taken under cw laser illumination, while the green hue represents the emitted THG signal. As opposed to the THG signal from the Si--SiO$_2$ substrate in Fig.2(b), and from the waveguide in Fig.2(c), the nanodisk provides a bright and distinct THG spot, as shown in Fig.2(d). Specifically, here, the THG signal from the nanodisk is stronger than that from the silicon substrate by a factor of 17. Given that, under the same conditions, smaller nanodisks provided an order of magnitude less intense THG, we are certain that the main contribution of THG in our observations come from the nanodisks with resonantly excited MD resonances. 
	
	In order to experimentally access different regimes of optical coupling between the resonant nanodisk and the waveguide, we expand our measurement scope by considering different gap sizes between the particles. In Table~1, a summary of the THG microscopy results is given for nine combinations of $D$ and $L$; here, THG from the nanostructures was normalized by the THG measured at the substrate. First, we quantify the difference between THG measured for the resonant nanodisks ($D=480$~nm) and non-resonant nanodisks ($D=430$ and 380~nm), with the latter providing up to $17.0/0.7=24$ times less THG yield, with a typical statistical error of $\pm$20\%.  Note that, for the smallest disks, a proper positioning of the pump beam on the nanodisks was not guaranteed, since the THG signal measured at the bare waveguide was on the same level ($\approx0.8$) with the THG from the nanodisks ($\approx0.7-0.9$). Second, a clear difference, by a factor of 5, is observed between the THG signals from the disks of the same diameter but different gap sizes, with a local maximum observed for a gap size of $L=185$~nm. Below, we numerically address this observation and show that the presence of a waveguide can enhance field localization in the nanodisk and constructively amend its nonlinear response.	In Supporting Information we provide measurement results for the orthogonal pump polarization; there, the waveguide itself demonstrates considerable, by a factor of 6, THG enhancement with respect to the substrate, which disrupts unambiguous interpretation.

\textbf{Numerical calculations} A full-wave simulation of the system under study was performed using commercially available Maxwell's equations solver, see Methods for details. In order to verify the strong dependence of the nonlinear-optical response of the nanodisk on its distance to the waveguide, and separate it from any other sources of THG, we calculated and integrated the local fields within the disk as a function of $L$. In order to mimic the third-order nonlinear response, we plot in Fig.3(a) with connected red squares the following quantity:
\begin{equation}
I_{\rm avg}=\int\limits_{\rm disk} |\mathbf{E}(\mathbf{r})|^6 dV,
\end{equation}
where $\mathbf{E}(\mathbf{r})$ is the calculated electric field strength distribution, which is integrated over the volume of the nanodisk only. 
Although this quantity is not straightforwardly connected to the measured THG signals, it gives us important insights about how the waveguide affects the local fields within the nanodisk. It is non-monotonic, and shows damped oscillations saturating at a level of about 0.4 for large $L$, where the waveguide is placed considerably far from the nanodisk. This means that, in agreement with our experimental observations, the waveguide, although non-resonant, can constructively amend the local fields within the nanodisk. The explanation of this behavior is given in Fig.3(b). We have calculated and plotted the central wavelength and full-width at half-maximum (FWHM) of the nanodisk's MD scattering resonance as a function of $L$, presented in Fig.3(b) with blue and black curves, respectively. It is very instructive that the FWHM of the resonance is strongly modulated by the waveguide, varying from 155~nm at $L\approx200$~nm to 250~nm at  $L\approx600$~nm.  By comparing the FWHM dependence on $L$ to that of the average local fields $I_{\rm avg}(L)$, given by the dashed red curve, one can conclude that the main contribution to the alternating local fields of the nanodisk is provided by the dependence of the MD resonance radiative decay constant.  We can therefore conclude that the highly non-monotonic dependence of the measured third harmonic signal as a function of the gap size represent an observation of efficient optical coupling between the nanodisk and the waveguide.

		\begin{table}[t]
		
		\begin{tabular}[b]{|p{3em}|p{5em}|p{5em}|p{5em}|p{5.8em}|}
			\hline
			\multicolumn{1}{|c|}{D}& \multicolumn{3}{c|}{Gap size L (nm)} & $\lambda_{\rm pump}-\lambda_{\rm MD}$ \\
			\multicolumn{1}{|c|}{(nm)} & \multicolumn{1}{c}{90} & \multicolumn{1}{c}{185} & \multicolumn{1}{c|}{315}  & \multicolumn{1}{c|}{(nm)}\\
			\hline
			380  & 0.7 & 0.9 
			& 0.9  &
			170 \\
			\hline
			430 & 1.0 & 2.4 & 1.4  & 90 \\
			\hline
			480 & 3.8  & 17.0  & 6.5   & 10 \\                                       \hline
		\end{tabular}
		\caption{\label{} Third harmonic generation from silicon nanodisks coupled to silicon waveguides normalized by the THG from the silicon substrate. The normalized signal from the waveguide without a nearby disk is 0.8. Typical relative signal errors are  $\pm$20\%.}
	\end{table}

	To conclude, we have performed nonlinear microscopy on resonant all-dielectric nanoparticles optically coupled to dielectric waveguides. Third harmonic generation from subwavelength silicon nanodisks is  enhanced by a factor of 25 with respect to bulk silicon when its fundamental magnetic dipolar mode is resonantly excited by femtosecond laser pulses. Moreover, the third harmonic signal is shown to significantly depend on the distance between the nanodisk and the waveguide, with the maximum detected modulation of up to 4.5. The presence of the waveguide was found to increase the third harmonic output from the nanodisk, which is explained by re-distribution of the local fields and reduction of radiative losses of the system, that result in a higher average third-order nonlinear polarization. We render this study an important step toward integration of resonant all-dielectric nanostructures on photonic chips.

\begin{figure}[t]
	\begin{overpic}[width=1\linewidth]{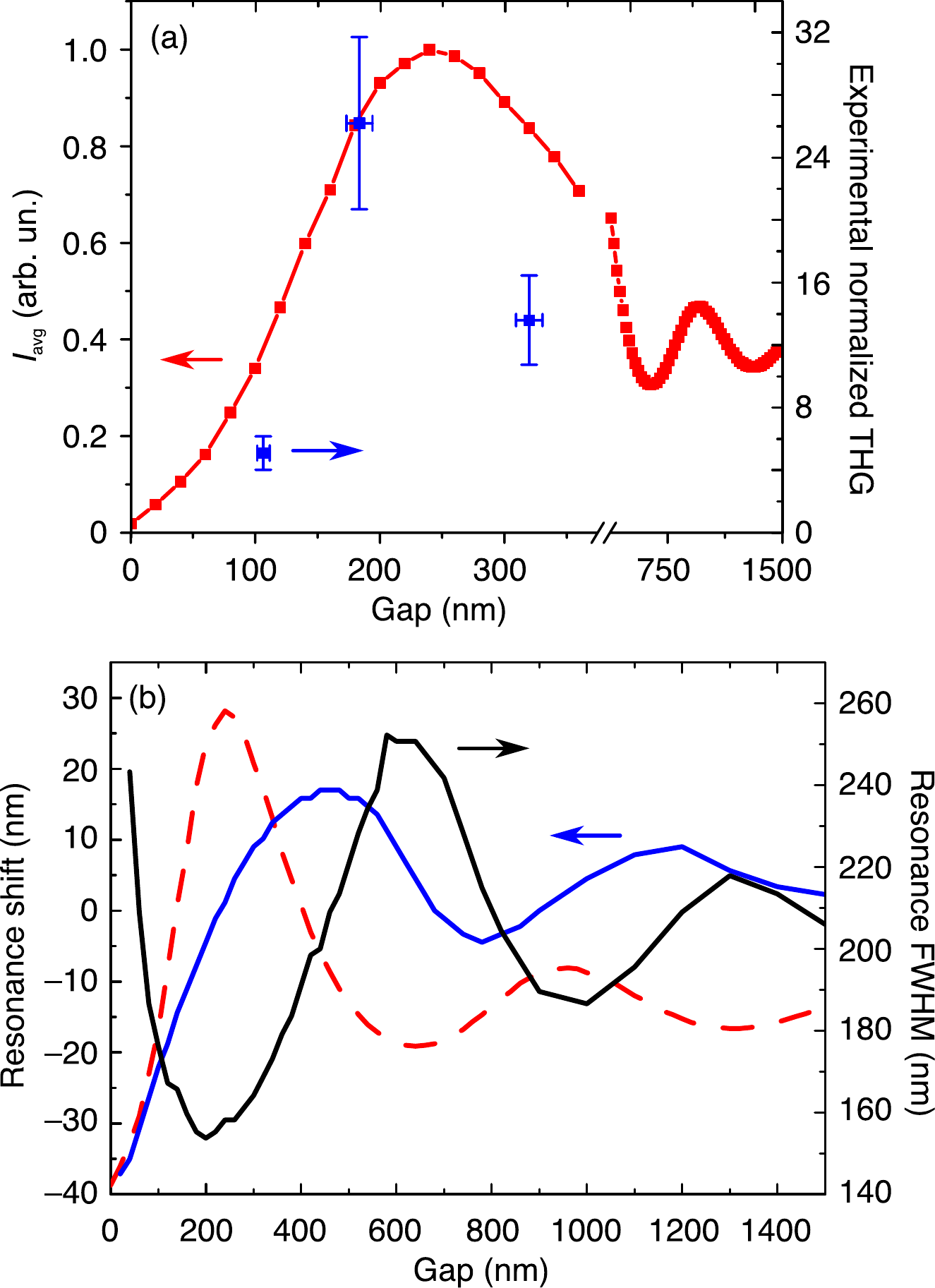}
	\end{overpic}
	\caption{\label{fig0} {(a) Experimental normalized THG enhancement from resonant ($D=480$~nm) nanodisks as a function of the gap between the nanodisk and the waveguide (blue squares) and normalized local electric fields, given by Eq.(1) (connected red  squares). (b) MD resonance parameters as a function of gap $L$, with its central position plotted in blue, and full width at half maximum plotted in black. The dashed line is the same as the red curve from panel (a).} 
	}
\end{figure}

	\section{Acknowledgments}
	
	Financial support from Russian Foundation for Basic Research (\#16-02-01092) is acknowledged.
	
	\section{Methods}

	\textbf{THG and confocal microscopy}
	For THG and laser confocal microscopy, we used an Olympus FluoView FV1000 laser scanning confocal microscope in combination with a femtosecond Er$^{3+}$ fiber laser (Avesta EFOA-100) with 120-fs-long pulses at a carrier photon energy of $\approx0.8$~eV and a repetition rate of 70 MHz. 
	The confocal microscope was equipped with a laser combiner that allowed multi-wavelength operation. As a result, THG and scanning confocal images were captured simultaneously as detected by two separate photomultipliers (PMT). Additionally, in order to avoid two-photon absorption at the PMT cathode, pump radiation was filtered out by a monochromator in front of one of the PMTs.
	In THG microscopy measurements, a femtosecond laser beam passed through a polarizer (Glan prism) and focused at the sample into a spot with a diameter of $\approx3$~$\mu$m by an aspheric lens with $NA=0.6$ and a focal length of 4~mm (ThorLabs C610TME-B). The polarizer orientation was chosen to be either perpendicular or parallel to the waveguide. The transmission radiation and third-harmonic frequency was collected by a 40x objective with $NA=0.9$.
	 
	\textbf{Calculations}
	For the numerical calculation of the THG and the local field distribution, we used the commercial software package Lumerical FDTD Solutions, in which simulation area was constrained by perfectly matched layer boundary conditions. 
	At the SiO$_{2}$ substrate side, this condition which mimics the response of a SiO$_{2}$ half space, was chosen to terminate the computational volume.
	A normally impinging broadband Gaussian beam was chosen as a light source. It was located in the SiO$_{2}$ substrate and focused into a waist of about 1.5~$\mu$m in diameter and divergence angle of $\approx5^\circ$. 
	The dimensions of the structure were as follows: the SiO$_{2}$ layer was 277 nm in thickness, the silicon waveguide was 280 nm in height and 445 nm in width, and the nanodisks were 280 nm in height and 380, 430 or 480 nm in diameter. 
	The gaps between silicon waveguide and silicon nanodisks were 90, 185 or 315 nm.
The following values of isotropic nonlinear susceptibilities were added to the default dielectric materials: $\chi^{(3)}_{\rm Si}\approx 2.45\cdot10^{-19} \frac{m^{2}}{V^{2}}$ 
and $\chi^{(3)}_{\rm SiO_{2}}\approx 2\cdot10^{-22} \frac{m^{2}}{V^{2}}$.
	
%



\begin{thebibliography}{23}%
\makeatletter
\providecommand \@ifxundefined [1]{%
 \@ifx{#1\undefined}
}%
\providecommand \@ifnum [1]{%
 \ifnum #1\expandafter \@firstoftwo
 \else \expandafter \@secondoftwo
 \fi
}%
\providecommand \@ifx [1]{%
 \ifx #1\expandafter \@firstoftwo
 \else \expandafter \@secondoftwo
 \fi
}%
\providecommand \natexlab [1]{#1}%
\providecommand \enquote  [1]{``#1''}%
\providecommand \bibnamefont  [1]{#1}%
\providecommand \bibfnamefont [1]{#1}%
\providecommand \citenamefont [1]{#1}%
\providecommand \href@noop [0]{\@secondoftwo}%
\providecommand \href [0]{\begingroup \@sanitize@url \@href}%
\providecommand \@href[1]{\@@startlink{#1}\@@href}%
\providecommand \@@href[1]{\endgroup#1\@@endlink}%
\providecommand \@sanitize@url [0]{\catcode `\\12\catcode `\$12\catcode
  `\&12\catcode `\#12\catcode `\^12\catcode `\_12\catcode `\%12\relax}%
\providecommand \@@startlink[1]{}%
\providecommand \@@endlink[0]{}%
\providecommand \url  [0]{\begingroup\@sanitize@url \@url }%
\providecommand \@url [1]{\endgroup\@href {#1}{\urlprefix }}%
\providecommand \urlprefix  [0]{URL }%
\providecommand \Eprint [0]{\href }%
\providecommand \doibase [0]{http://dx.doi.org/}%
\providecommand \selectlanguage [0]{\@gobble}%
\providecommand \bibinfo  [0]{\@secondoftwo}%
\providecommand \bibfield  [0]{\@secondoftwo}%
\providecommand \translation [1]{[#1]}%
\providecommand \BibitemOpen [0]{}%
\providecommand \bibitemStop [0]{}%
\providecommand \bibitemNoStop [0]{.\EOS\space}%
\providecommand \EOS [0]{\spacefactor3000\relax}%
\providecommand \BibitemShut  [1]{\csname bibitem#1\endcsname}%
\let\auto@bib@innerbib\@empty
\bibitem [{\citenamefont {Atwater}\ and\ \citenamefont
  {Polman}(2010)}]{Atwater2010}%
  \BibitemOpen
  \bibfield  {author} {\bibinfo {author} {\bibfnamefont {H.}~\bibnamefont
  {Atwater}}\ and\ \bibinfo {author} {\bibfnamefont {A.}~\bibnamefont
  {Polman}},\ }\href {http://www.worldscibooks.com/environsci/7848.html}
  {\bibfield  {journal} {\bibinfo  {journal} {Nature Mater.}\ }\textbf
  {\bibinfo {volume} {9}},\ \bibinfo {pages} {205} (\bibinfo {year}
  {2010})}\BibitemShut {NoStop}%
\bibitem [{\citenamefont {Jain}\ \emph {et~al.}(2008)\citenamefont {Jain},
  \citenamefont {Huang}, \citenamefont {El-Sayed},\ and\ \citenamefont
  {El-Sayed}}]{Jain2008}%
  \BibitemOpen
  \bibfield  {author} {\bibinfo {author} {\bibfnamefont {P.~K.}\ \bibnamefont
  {Jain}}, \bibinfo {author} {\bibfnamefont {X.}~\bibnamefont {Huang}},
  \bibinfo {author} {\bibfnamefont {I.~H.}\ \bibnamefont {El-Sayed}}, \ and\
  \bibinfo {author} {\bibfnamefont {M.~A.}\ \bibnamefont {El-Sayed}},\ }\href
  {\doibase 10.1021/ar7002804} {\bibfield  {journal} {\bibinfo  {journal} {Acc.
  Chem. Res.}\ }\textbf {\bibinfo {volume} {41}},\ \bibinfo {pages} {1578}
  (\bibinfo {year} {2008})}\BibitemShut {NoStop}%
\bibitem [{\citenamefont {Jain}\ \emph {et~al.}(2006)\citenamefont {Jain},
  \citenamefont {Lee}, \citenamefont {El-Sayed},\ and\ \citenamefont
  {El-Sayed}}]{Jain2006}%
  \BibitemOpen
  \bibfield  {author} {\bibinfo {author} {\bibfnamefont {P.~K.}\ \bibnamefont
  {Jain}}, \bibinfo {author} {\bibfnamefont {K.~S.}\ \bibnamefont {Lee}},
  \bibinfo {author} {\bibfnamefont {I.~H.}\ \bibnamefont {El-Sayed}}, \ and\
  \bibinfo {author} {\bibfnamefont {M.~A.}\ \bibnamefont {El-Sayed}},\ }\href
  {\doibase 10.1021/jp057170o} {\bibfield  {journal} {\bibinfo  {journal} {J.
  Phys. Chem. B}\ }\textbf {\bibinfo {volume} {110}},\ \bibinfo {pages} {7238}
  (\bibinfo {year} {2006})}\BibitemShut {NoStop}%
\bibitem [{\citenamefont {Kelly}\ \emph {et~al.}(2003)\citenamefont {Kelly},
  \citenamefont {Coronado}, \citenamefont {Zhao},\ and\ \citenamefont
  {Schatz}}]{Kelly2003}%
  \BibitemOpen
  \bibfield  {author} {\bibinfo {author} {\bibfnamefont {K.~L.}\ \bibnamefont
  {Kelly}}, \bibinfo {author} {\bibfnamefont {E.}~\bibnamefont {Coronado}},
  \bibinfo {author} {\bibfnamefont {L.~L.}\ \bibnamefont {Zhao}}, \ and\
  \bibinfo {author} {\bibfnamefont {G.~C.}\ \bibnamefont {Schatz}},\ }\href
  {\doibase 10.1021/jp026731y} {\bibfield  {journal} {\bibinfo  {journal} {J.
  Phys. Chem. B}\ }\textbf {\bibinfo {volume} {107}},\ \bibinfo {pages} {668}
  (\bibinfo {year} {2003})}\BibitemShut {NoStop}%
\bibitem [{\citenamefont {Rechberger}\ \emph {et~al.}(2003)\citenamefont
  {Rechberger}, \citenamefont {Hohenau}, \citenamefont {Leitner}, \citenamefont
  {Krenn}, \citenamefont {Lamprecht},\ and\ \citenamefont
  {Aussenegg}}]{Rechberger2003}%
  \BibitemOpen
  \bibfield  {author} {\bibinfo {author} {\bibfnamefont {W.}~\bibnamefont
  {Rechberger}}, \bibinfo {author} {\bibfnamefont {A.}~\bibnamefont {Hohenau}},
  \bibinfo {author} {\bibfnamefont {A.}~\bibnamefont {Leitner}}, \bibinfo
  {author} {\bibfnamefont {J.~R.}\ \bibnamefont {Krenn}}, \bibinfo {author}
  {\bibfnamefont {B.}~\bibnamefont {Lamprecht}}, \ and\ \bibinfo {author}
  {\bibfnamefont {F.~R.}\ \bibnamefont {Aussenegg}},\ }\href {\doibase
  10.1016/S0030-4018(03)01357-9} {\bibfield  {journal} {\bibinfo  {journal}
  {Opt. Commun.}\ }\textbf {\bibinfo {volume} {220}},\ \bibinfo {pages} {137}
  (\bibinfo {year} {2003})}\BibitemShut {NoStop}%
\bibitem [{\citenamefont {Halas}\ \emph {et~al.}(2011)\citenamefont {Halas},
  \citenamefont {Lal}, \citenamefont {Chang}, \citenamefont {Link},\ and\
  \citenamefont {Nordlander}}]{Halas2011}%
  \BibitemOpen
  \bibfield  {author} {\bibinfo {author} {\bibfnamefont {N.~J.}\ \bibnamefont
  {Halas}}, \bibinfo {author} {\bibfnamefont {S.}~\bibnamefont {Lal}}, \bibinfo
  {author} {\bibfnamefont {W.~S.}\ \bibnamefont {Chang}}, \bibinfo {author}
  {\bibfnamefont {S.}~\bibnamefont {Link}}, \ and\ \bibinfo {author}
  {\bibfnamefont {P.}~\bibnamefont {Nordlander}},\ }\href {\doibase
  10.1021/cr200061k} {\bibfield  {journal} {\bibinfo  {journal} {Chem. Rev.}\
  }\textbf {\bibinfo {volume} {111}},\ \bibinfo {pages} {3913} (\bibinfo {year}
  {2011})}\BibitemShut {NoStop}%
\bibitem [{\citenamefont {Liu}\ \emph {et~al.}(2011)\citenamefont {Liu},
  \citenamefont {Hentschel}, \citenamefont {Weiss}, \citenamefont
  {Alivisatos},\ and\ \citenamefont {Giessen}}]{Liu2011}%
  \BibitemOpen
  \bibfield  {author} {\bibinfo {author} {\bibfnamefont {N.}~\bibnamefont
  {Liu}}, \bibinfo {author} {\bibfnamefont {M.}~\bibnamefont {Hentschel}},
  \bibinfo {author} {\bibfnamefont {T.}~\bibnamefont {Weiss}}, \bibinfo
  {author} {\bibfnamefont {A.~P.}\ \bibnamefont {Alivisatos}}, \ and\ \bibinfo
  {author} {\bibfnamefont {H.}~\bibnamefont {Giessen}},\ }\href {\doibase
  10.1126/science.1199958} {\bibfield  {journal} {\bibinfo  {journal}
  {Science}\ }\textbf {\bibinfo {volume} {332}},\ \bibinfo {pages} {1407}
  (\bibinfo {year} {2011})}\BibitemShut {NoStop}%
\bibitem [{\citenamefont {Maier}\ \emph {et~al.}(2003)\citenamefont {Maier},
  \citenamefont {Kik}, \citenamefont {Atwater}, \citenamefont {Meltzer},
  \citenamefont {Harel}, \citenamefont {Koel},\ and\ \citenamefont
  {Requicha}}]{Maier2003}%
  \BibitemOpen
  \bibfield  {author} {\bibinfo {author} {\bibfnamefont {S.~A.}\ \bibnamefont
  {Maier}}, \bibinfo {author} {\bibfnamefont {P.~G.}\ \bibnamefont {Kik}},
  \bibinfo {author} {\bibfnamefont {H.~A.}\ \bibnamefont {Atwater}}, \bibinfo
  {author} {\bibfnamefont {S.}~\bibnamefont {Meltzer}}, \bibinfo {author}
  {\bibfnamefont {E.}~\bibnamefont {Harel}}, \bibinfo {author} {\bibfnamefont
  {B.~E.}\ \bibnamefont {Koel}}, \ and\ \bibinfo {author} {\bibfnamefont
  {A.~A.}\ \bibnamefont {Requicha}},\ }\href {\doibase 10.1038/nmat852}
  {\bibfield  {journal} {\bibinfo  {journal} {Nature Mater.}\ }\textbf
  {\bibinfo {volume} {2}},\ \bibinfo {pages} {229} (\bibinfo {year}
  {2003})}\BibitemShut {NoStop}%
\bibitem [{\citenamefont {Jain}\ \emph {et~al.}(2007)\citenamefont {Jain},
  \citenamefont {Huang},\ and\ \citenamefont {El-Sayed}}]{Jain2007}%
  \BibitemOpen
  \bibfield  {author} {\bibinfo {author} {\bibfnamefont {P.~K.}\ \bibnamefont
  {Jain}}, \bibinfo {author} {\bibfnamefont {W.}~\bibnamefont {Huang}}, \ and\
  \bibinfo {author} {\bibfnamefont {M.~A.}\ \bibnamefont {El-Sayed}},\ }\href
  {\doibase 10.1021/nl071008a} {\bibfield  {journal} {\bibinfo  {journal} {Nano
  Lett.}\ }\textbf {\bibinfo {volume} {7}},\ \bibinfo {pages} {2080} (\bibinfo
  {year} {2007})}\BibitemShut {NoStop}%
\bibitem [{\citenamefont {Teperik}\ \emph {et~al.}(2013)\citenamefont
  {Teperik}, \citenamefont {Nordlander}, \citenamefont {Aizpurua},\ and\
  \citenamefont {Borisov}}]{Teperik2013}%
  \BibitemOpen
  \bibfield  {author} {\bibinfo {author} {\bibfnamefont {T.~V.}\ \bibnamefont
  {Teperik}}, \bibinfo {author} {\bibfnamefont {P.}~\bibnamefont {Nordlander}},
  \bibinfo {author} {\bibfnamefont {J.}~\bibnamefont {Aizpurua}}, \ and\
  \bibinfo {author} {\bibfnamefont {A.~G.}\ \bibnamefont {Borisov}},\ }\href
  {\doibase 10.1103/PhysRevLett.110.263901} {\bibfield  {journal} {\bibinfo
  {journal} {Phys. Rev. Lett.}\ }\textbf {\bibinfo {volume} {110}},\ \bibinfo
  {pages} {263901} (\bibinfo {year} {2013})}\BibitemShut {NoStop}%
\bibitem [{\citenamefont {Shcherbakov}\ \emph {et~al.}(2015)\citenamefont
  {Shcherbakov}, \citenamefont {Shorokhov}, \citenamefont {Neshev},
  \citenamefont {Hopkins}, \citenamefont {Staude}, \citenamefont
  {Melik-Gaykazyan}, \citenamefont {Ezhov}, \citenamefont {Miroshnichenko},
  \citenamefont {Brener}, \citenamefont {Fedyanin},\ and\ \citenamefont
  {Kivshar}}]{Shcherbakov2015}%
  \BibitemOpen
  \bibfield  {author} {\bibinfo {author} {\bibfnamefont {M.~R.}\ \bibnamefont
  {Shcherbakov}}, \bibinfo {author} {\bibfnamefont {A.~S.}\ \bibnamefont
  {Shorokhov}}, \bibinfo {author} {\bibfnamefont {D.~N.}\ \bibnamefont
  {Neshev}}, \bibinfo {author} {\bibfnamefont {B.}~\bibnamefont {Hopkins}},
  \bibinfo {author} {\bibfnamefont {I.}~\bibnamefont {Staude}}, \bibinfo
  {author} {\bibfnamefont {E.~V.}\ \bibnamefont {Melik-Gaykazyan}}, \bibinfo
  {author} {\bibfnamefont {A.~A.}\ \bibnamefont {Ezhov}}, \bibinfo {author}
  {\bibfnamefont {A.~E.}\ \bibnamefont {Miroshnichenko}}, \bibinfo {author}
  {\bibfnamefont {I.}~\bibnamefont {Brener}}, \bibinfo {author} {\bibfnamefont
  {A.~A.}\ \bibnamefont {Fedyanin}}, \ and\ \bibinfo {author} {\bibfnamefont
  {Y.~S.}\ \bibnamefont {Kivshar}},\ }\href {\doibase
  10.1021/acsphotonics.5b00065} {\bibfield  {journal} {\bibinfo  {journal} {ACS
  Photonics}\ }\textbf {\bibinfo {volume} {2}},\ \bibinfo {pages} {578}
  (\bibinfo {year} {2015})}\BibitemShut {NoStop}%
\bibitem [{\citenamefont {Kuznetsov}\ \emph {et~al.}(2016)\citenamefont
  {Kuznetsov}, \citenamefont {Miroshnichenko}, \citenamefont {Brongersma},
  \citenamefont {Kivshar},\ and\ \citenamefont {Lukyanchuk}}]{Kuznetsov2016a}%
  \BibitemOpen
  \bibfield  {author} {\bibinfo {author} {\bibfnamefont {A.~I.}\ \bibnamefont
  {Kuznetsov}}, \bibinfo {author} {\bibfnamefont {A.~E.}\ \bibnamefont
  {Miroshnichenko}}, \bibinfo {author} {\bibfnamefont {M.~L.}\ \bibnamefont
  {Brongersma}}, \bibinfo {author} {\bibfnamefont {Y.~S.}\ \bibnamefont
  {Kivshar}}, \ and\ \bibinfo {author} {\bibfnamefont {B.}~\bibnamefont
  {Lukyanchuk}},\ }\href {\doibase 10.1126/science.aag2472} {\bibfield
  {journal} {\bibinfo  {journal} {Science}\ }\textbf {\bibinfo {volume}
  {354}},\ \bibinfo {pages} {aag2472} (\bibinfo {year} {2016})}\BibitemShut
  {NoStop}%
\bibitem [{\citenamefont {Kuznetsov}\ \emph {et~al.}(2012)\citenamefont
  {Kuznetsov}, \citenamefont {Miroshnichenko}, \citenamefont {Fu},
  \citenamefont {Zhang},\ and\ \citenamefont {Luk'yanchuk}}]{Kuznetsov2012}%
  \BibitemOpen
  \bibfield  {author} {\bibinfo {author} {\bibfnamefont {A.~I.}\ \bibnamefont
  {Kuznetsov}}, \bibinfo {author} {\bibfnamefont {A.~E.}\ \bibnamefont
  {Miroshnichenko}}, \bibinfo {author} {\bibfnamefont {Y.~H.}\ \bibnamefont
  {Fu}}, \bibinfo {author} {\bibfnamefont {J.}~\bibnamefont {Zhang}}, \ and\
  \bibinfo {author} {\bibfnamefont {B.}~\bibnamefont {Luk'yanchuk}},\ }\href
  {\doibase 10.1038/srep00492} {\bibfield  {journal} {\bibinfo  {journal} {Sci.
  Rep.}\ }\textbf {\bibinfo {volume} {2}},\ \bibinfo {pages} {492} (\bibinfo
  {year} {2012})}\BibitemShut {NoStop}%
\bibitem [{\citenamefont {Evlyukhin}\ \emph {et~al.}(2012)\citenamefont
  {Evlyukhin}, \citenamefont {Novikov}, \citenamefont {Zywietz}, \citenamefont
  {Eriksen}, \citenamefont {Reinhardt}, \citenamefont {Bozhevolnyi},\ and\
  \citenamefont {Chichkov}}]{Evlyukhin2012}%
  \BibitemOpen
  \bibfield  {author} {\bibinfo {author} {\bibfnamefont {A.~B.}\ \bibnamefont
  {Evlyukhin}}, \bibinfo {author} {\bibfnamefont {S.~M.}\ \bibnamefont
  {Novikov}}, \bibinfo {author} {\bibfnamefont {U.}~\bibnamefont {Zywietz}},
  \bibinfo {author} {\bibfnamefont {R.~L.}\ \bibnamefont {Eriksen}}, \bibinfo
  {author} {\bibfnamefont {C.}~\bibnamefont {Reinhardt}}, \bibinfo {author}
  {\bibfnamefont {S.~I.}\ \bibnamefont {Bozhevolnyi}}, \ and\ \bibinfo {author}
  {\bibfnamefont {B.~N.}\ \bibnamefont {Chichkov}},\ }\href {\doibase
  10.1021/nl301594s} {\bibfield  {journal} {\bibinfo  {journal} {Nano Lett.}\
  }\textbf {\bibinfo {volume} {12}},\ \bibinfo {pages} {3749} (\bibinfo {year}
  {2012})}\BibitemShut {NoStop}%
\bibitem [{\citenamefont {Bakker}\ \emph {et~al.}(2015)\citenamefont {Bakker},
  \citenamefont {Permyakov}, \citenamefont {Yu}, \citenamefont {Markovich},
  \citenamefont {Paniagua-Dom{\'{i}}nguez}, \citenamefont {Gonzaga},
  \citenamefont {Samusev}, \citenamefont {Kivshar}, \citenamefont
  {Lukyanchuk},\ and\ \citenamefont {Kuznetsov}}]{Bakker2015}%
  \BibitemOpen
  \bibfield  {author} {\bibinfo {author} {\bibfnamefont {R.~M.}\ \bibnamefont
  {Bakker}}, \bibinfo {author} {\bibfnamefont {D.}~\bibnamefont {Permyakov}},
  \bibinfo {author} {\bibfnamefont {Y.~F.}\ \bibnamefont {Yu}}, \bibinfo
  {author} {\bibfnamefont {D.}~\bibnamefont {Markovich}}, \bibinfo {author}
  {\bibfnamefont {R.}~\bibnamefont {Paniagua-Dom{\'{i}}nguez}}, \bibinfo
  {author} {\bibfnamefont {L.}~\bibnamefont {Gonzaga}}, \bibinfo {author}
  {\bibfnamefont {A.}~\bibnamefont {Samusev}}, \bibinfo {author} {\bibfnamefont
  {Y.}~\bibnamefont {Kivshar}}, \bibinfo {author} {\bibfnamefont
  {B.}~\bibnamefont {Lukyanchuk}}, \ and\ \bibinfo {author} {\bibfnamefont
  {A.~I.}\ \bibnamefont {Kuznetsov}},\ }\href {\doibase
  10.1021/acs.nanolett.5b00128} {\bibfield  {journal} {\bibinfo  {journal}
  {Nano Lett.}\ }\textbf {\bibinfo {volume} {15}},\ \bibinfo {pages} {2137}
  (\bibinfo {year} {2015})}\BibitemShut {NoStop}%
\bibitem [{\citenamefont {Hopkins}\ \emph {et~al.}(2013)\citenamefont
  {Hopkins}, \citenamefont {Poddubny}, \citenamefont {Miroshnichenko},\ and\
  \citenamefont {Kivshar}}]{Hopkins2013}%
  \BibitemOpen
  \bibfield  {author} {\bibinfo {author} {\bibfnamefont {B.}~\bibnamefont
  {Hopkins}}, \bibinfo {author} {\bibfnamefont {A.~N.}\ \bibnamefont
  {Poddubny}}, \bibinfo {author} {\bibfnamefont {A.~E.}\ \bibnamefont
  {Miroshnichenko}}, \ and\ \bibinfo {author} {\bibfnamefont {Y.~S.}\
  \bibnamefont {Kivshar}},\ }\href {\doibase 10.1103/PhysRevA.88.053819}
  {\bibfield  {journal} {\bibinfo  {journal} {Phys. Rev. A}\ }\textbf {\bibinfo
  {volume} {88}},\ \bibinfo {pages} {053819} (\bibinfo {year}
  {2013})}\BibitemShut {NoStop}%
\bibitem [{\citenamefont {Shorokhov}\ \emph {et~al.}(2016)\citenamefont
  {Shorokhov}, \citenamefont {Melik-Gaykazyan}, \citenamefont {Smirnova},
  \citenamefont {Hopkins}, \citenamefont {Chong}, \citenamefont {Choi},
  \citenamefont {Shcherbakov}, \citenamefont {Miroshnichenko}, \citenamefont
  {Neshev}, \citenamefont {Fedyanin},\ and\ \citenamefont
  {Kivshar}}]{Shorokhov2016}%
  \BibitemOpen
  \bibfield  {author} {\bibinfo {author} {\bibfnamefont {A.~S.}\ \bibnamefont
  {Shorokhov}}, \bibinfo {author} {\bibfnamefont {E.~V.}\ \bibnamefont
  {Melik-Gaykazyan}}, \bibinfo {author} {\bibfnamefont {D.~A.}\ \bibnamefont
  {Smirnova}}, \bibinfo {author} {\bibfnamefont {B.}~\bibnamefont {Hopkins}},
  \bibinfo {author} {\bibfnamefont {K.~E.}\ \bibnamefont {Chong}}, \bibinfo
  {author} {\bibfnamefont {D.~Y.}\ \bibnamefont {Choi}}, \bibinfo {author}
  {\bibfnamefont {M.~R.}\ \bibnamefont {Shcherbakov}}, \bibinfo {author}
  {\bibfnamefont {A.~E.}\ \bibnamefont {Miroshnichenko}}, \bibinfo {author}
  {\bibfnamefont {D.~N.}\ \bibnamefont {Neshev}}, \bibinfo {author}
  {\bibfnamefont {A.~A.}\ \bibnamefont {Fedyanin}}, \ and\ \bibinfo {author}
  {\bibfnamefont {Y.~S.}\ \bibnamefont {Kivshar}},\ }\href {\doibase
  10.1021/acs.nanolett.6b01249} {\bibfield  {journal} {\bibinfo  {journal}
  {Nano Lett.}\ }\textbf {\bibinfo {volume} {16}},\ \bibinfo {pages} {4857}
  (\bibinfo {year} {2016})}\BibitemShut {NoStop}%
\bibitem [{\citenamefont {Shcherbakov}\ \emph {et~al.}(2014)\citenamefont
  {Shcherbakov}, \citenamefont {Neshev}, \citenamefont {Hopkins}, \citenamefont
  {Shorokhov}, \citenamefont {Staude}, \citenamefont {Melik-Gaykazyan},
  \citenamefont {Decker}, \citenamefont {Ezhov}, \citenamefont
  {Miroshnichenko}, \citenamefont {Brener}, \citenamefont {Fedyanin},\ and\
  \citenamefont {Kivshar}}]{Shcherbakov2014}%
  \BibitemOpen
  \bibfield  {author} {\bibinfo {author} {\bibfnamefont {M.~R.}\ \bibnamefont
  {Shcherbakov}}, \bibinfo {author} {\bibfnamefont {D.~N.}\ \bibnamefont
  {Neshev}}, \bibinfo {author} {\bibfnamefont {B.}~\bibnamefont {Hopkins}},
  \bibinfo {author} {\bibfnamefont {A.~S.}\ \bibnamefont {Shorokhov}}, \bibinfo
  {author} {\bibfnamefont {I.}~\bibnamefont {Staude}}, \bibinfo {author}
  {\bibfnamefont {E.~V.}\ \bibnamefont {Melik-Gaykazyan}}, \bibinfo {author}
  {\bibfnamefont {M.}~\bibnamefont {Decker}}, \bibinfo {author} {\bibfnamefont
  {A.~A.}\ \bibnamefont {Ezhov}}, \bibinfo {author} {\bibfnamefont {A.~E.}\
  \bibnamefont {Miroshnichenko}}, \bibinfo {author} {\bibfnamefont
  {I.}~\bibnamefont {Brener}}, \bibinfo {author} {\bibfnamefont {A.~A.}\
  \bibnamefont {Fedyanin}}, \ and\ \bibinfo {author} {\bibfnamefont {Y.~S.}\
  \bibnamefont {Kivshar}},\ }\href
  {http://pubs.acs.org/doi/abs/10.1021/nl503029j} {\bibfield  {journal}
  {\bibinfo  {journal} {Nano Lett.}\ }\textbf {\bibinfo {volume} {14}},\
  \bibinfo {pages} {6488} (\bibinfo {year} {2014})}\BibitemShut {NoStop}%
\bibitem [{\citenamefont {Yang}\ \emph {et~al.}(2015)\citenamefont {Yang},
  \citenamefont {Wang}, \citenamefont {Boulesbaa}, \citenamefont {Kravchenko},
  \citenamefont {Briggs}, \citenamefont {Puretzky}, \citenamefont {Geohegan},\
  and\ \citenamefont {Valentine}}]{Yang2015b}%
  \BibitemOpen
  \bibfield  {author} {\bibinfo {author} {\bibfnamefont {Y.}~\bibnamefont
  {Yang}}, \bibinfo {author} {\bibfnamefont {W.}~\bibnamefont {Wang}}, \bibinfo
  {author} {\bibfnamefont {A.}~\bibnamefont {Boulesbaa}}, \bibinfo {author}
  {\bibfnamefont {I.~I.}\ \bibnamefont {Kravchenko}}, \bibinfo {author}
  {\bibfnamefont {D.~P.}\ \bibnamefont {Briggs}}, \bibinfo {author}
  {\bibfnamefont {A.}~\bibnamefont {Puretzky}}, \bibinfo {author}
  {\bibfnamefont {D.}~\bibnamefont {Geohegan}}, \ and\ \bibinfo {author}
  {\bibfnamefont {J.}~\bibnamefont {Valentine}},\ }\href {\doibase
  10.1021/acs.nanolett.5b02802} {\bibfield  {journal} {\bibinfo  {journal}
  {Nano Lett.}\ }\textbf {\bibinfo {volume} {15}},\ \bibinfo {pages} {7388}
  (\bibinfo {year} {2015})}\BibitemShut {NoStop}%
\bibitem [{\citenamefont {Liu}\ \emph {et~al.}(2016)\citenamefont {Liu},
  \citenamefont {Sinclair}, \citenamefont {Saravi}, \citenamefont {Keeler},
  \citenamefont {Yang}, \citenamefont {Reno}, \citenamefont {Peake},
  \citenamefont {Setzpfandt}, \citenamefont {Staude}, \citenamefont {Pertsch},\
  and\ \citenamefont {Brener}}]{Liu2016b}%
  \BibitemOpen
  \bibfield  {author} {\bibinfo {author} {\bibfnamefont {S.}~\bibnamefont
  {Liu}}, \bibinfo {author} {\bibfnamefont {M.~B.}\ \bibnamefont {Sinclair}},
  \bibinfo {author} {\bibfnamefont {S.}~\bibnamefont {Saravi}}, \bibinfo
  {author} {\bibfnamefont {G.~A.}\ \bibnamefont {Keeler}}, \bibinfo {author}
  {\bibfnamefont {Y.}~\bibnamefont {Yang}}, \bibinfo {author} {\bibfnamefont
  {J.}~\bibnamefont {Reno}}, \bibinfo {author} {\bibfnamefont {G.~M.}\
  \bibnamefont {Peake}}, \bibinfo {author} {\bibfnamefont {F.}~\bibnamefont
  {Setzpfandt}}, \bibinfo {author} {\bibfnamefont {I.}~\bibnamefont {Staude}},
  \bibinfo {author} {\bibfnamefont {T.}~\bibnamefont {Pertsch}}, \ and\
  \bibinfo {author} {\bibfnamefont {I.}~\bibnamefont {Brener}},\ }\href
  {\doibase 10.1021/acs.nanolett.6b01816} {\bibfield  {journal} {\bibinfo
  {journal} {Nano Lett.}\ }\textbf {\bibinfo {volume} {16}},\ \bibinfo {pages}
  {5426} (\bibinfo {year} {2016})}\BibitemShut {NoStop}%
\bibitem [{\citenamefont {Staude}\ and\ \citenamefont
  {Schilling}(2017)}]{Staude2017}%
  \BibitemOpen
  \bibfield  {author} {\bibinfo {author} {\bibfnamefont {I.}~\bibnamefont
  {Staude}}\ and\ \bibinfo {author} {\bibfnamefont {J.}~\bibnamefont
  {Schilling}},\ }\href {\doibase 10.1038/nphoton.2017.39} {\bibfield
  {journal} {\bibinfo  {journal} {Nature Photon.}\ }\textbf {\bibinfo {volume}
  {11}},\ \bibinfo {pages} {274} (\bibinfo {year} {2017})}\BibitemShut
  {NoStop}%
\bibitem [{\citenamefont {Sidiropoulos}\ \emph {et~al.}(2014)\citenamefont
  {Sidiropoulos}, \citenamefont {Nielsen}, \citenamefont {Roschuk},
  \citenamefont {Zayats}, \citenamefont {Maier},\ and\ \citenamefont
  {Oulton}}]{Sidiropoulos2014}%
  \BibitemOpen
  \bibfield  {author} {\bibinfo {author} {\bibfnamefont {T.~P.}\ \bibnamefont
  {Sidiropoulos}}, \bibinfo {author} {\bibfnamefont {M.~P.}\ \bibnamefont
  {Nielsen}}, \bibinfo {author} {\bibfnamefont {T.~R.}\ \bibnamefont
  {Roschuk}}, \bibinfo {author} {\bibfnamefont {A.~V.}\ \bibnamefont {Zayats}},
  \bibinfo {author} {\bibfnamefont {S.~A.}\ \bibnamefont {Maier}}, \ and\
  \bibinfo {author} {\bibfnamefont {R.~F.}\ \bibnamefont {Oulton}},\ }\href
  {\doibase 10.1021/ph5002796} {\bibfield  {journal} {\bibinfo  {journal} {ACS
  Photonics}\ }\textbf {\bibinfo {volume} {1}},\ \bibinfo {pages} {912}
  (\bibinfo {year} {2014})}\BibitemShut {NoStop}%
\bibitem [{\citenamefont {Vercruysse}\ \emph {et~al.}(2017)\citenamefont
  {Vercruysse}, \citenamefont {Neutens}, \citenamefont {Lagae}, \citenamefont
  {Verellen},\ and\ \citenamefont {{Van Dorpe}}}]{Vercruysse2017a}%
  \BibitemOpen
  \bibfield  {author} {\bibinfo {author} {\bibfnamefont {D.}~\bibnamefont
  {Vercruysse}}, \bibinfo {author} {\bibfnamefont {P.}~\bibnamefont {Neutens}},
  \bibinfo {author} {\bibfnamefont {L.}~\bibnamefont {Lagae}}, \bibinfo
  {author} {\bibfnamefont {N.}~\bibnamefont {Verellen}}, \ and\ \bibinfo
  {author} {\bibfnamefont {P.}~\bibnamefont {{Van Dorpe}}},\ }\href {\doibase
  10.1021/acsphotonics.7b00038} {\bibfield  {journal} {\bibinfo  {journal} {ACS
  Photonics}\ }\textbf {\bibinfo {volume} {4}},\ \bibinfo {pages} {1398}
  (\bibinfo {year} {2017})}\BibitemShut {NoStop}%
\end{thebibliography}
\end{document}